\documentclass[12pt]{iopart}

\usepackage{subfig}
\expandafter\let\csname equation*\endcsname=\relax
\expandafter\let\csname endequation*\endcsname=\relax
\usepackage{amsmath}
\usepackage{graphicx}

\usepackage{iopams,cite}  
\begin{document}

\title[Ground-state Bethe root densities and quantum phase transitions]{Ground-state Bethe root densities and quantum phase transitions}

\author{Jon Links and Ian Marquette}

\address{Centre for Mathematical Physics, School of Mathematics and Physics, \\ The University of Queensland, Brisbane, QLD 4072, Australia}

\eads{\mailto{jrl@maths.uq.edu.au},\,\,\mailto{i.marquette@uq.edu.au}}

\begin{abstract}
Exactly solvable models provide a unique method, via qualitative changes in the distribution of the ground-state roots of the Bethe Ansatz equations, to identify quantum phase transitions. Here we expand on this approach, in a quantitative manner, for two models of Bose--Einstein condensates. The first model deals with the interconversion of bosonic atoms and molecules. The second is the two-site Bose--Hubbard model, widely used to describe tunneling phenomena in Bose--Einstein condensates. For these systems we calculate the ground-state root density. This facilitates the determination of analytic forms for the ground-state energy, and associated correlation functions through the Hellmann--Feynman theorem. These calculations provide a clear  identification of the quantum phase transition in each model. For the first model we obtain an expression for the molecular fraction expectation value. For the two-site Bose--Hubbard model we find that there is a simple characterisation of condensate fragmentation.    
\end{abstract}

\maketitle

\section{Introduction}

In \cite{Rub12} Rubeni et al. studied quantum phase transitions in two bosonic models related to Bose--Einstein condensation, from the perspective of their Bethe Ansatz solutions. One model deals with the interconversion of bosonic atoms and molecules \cite{Dun07,Zhou07,Zhou08}. The other is the two-site Bose--Hubbard model, widely used to describe tunneling phenomena in Bose--Einstein condensates \cite{Mil97,Cir98,Leg01,Kohl02,Zhou03,Pan05} and which continues to be the subject of extensive study, e.g. \cite{Perez10,Polls10,zngo10,Buo12,Sim12,Jez13,Gra14,Sak14}. For these systems the quantum phase transition points are analogs of fixed-point bifurcations in a corresponding classical system. Crossing through a bifurcation leads to an abrupt change in the dynamical behaviour. Experimental observation of this property has been reported for a system modelled by the two-site Bose--Hubbard Hamiltonian \cite{zngo10}, raising the potential to probe quantum systems at the macroscopic level. This feature has also been reported in a photonic context \cite{a13}.   

By numerically solving the Bethe ansatz equations for the ground state, it was found in \cite{Rub12} that there is a sharp change in the character of the root distribution in the complex plane around a particular coupling value. Through complementary computations of entanglement, fidelity, and the energy gap, it was identified that the change in the root distribution coincides with a quantum phase transition. Similar correspondences have also been witnessed in other models admitting exact Bethe ansatz solutions \cite{Dun10,Rom10,Ler11,Mar12,Ler13,Mar13,Van14}, which have in common that their solutions are of the {\it Richardson--Gaudin} form. In some literature this form is also referred to as Bethe Ansatz equations in the quasi-classical limit. See \cite{lil14} and references therein for a summary of this latter point.

The goal of the present research is to provide an enhanced quantitative study of the two models considered in \cite{Rub12}.  We will approach this from both an analytic and a numerical viewpoint. The techniques implemented are quite general and can be applied to other models with Bethe Ansatz equations of the Richardson--Gaudin form, which are the following class of coupled nonlinear algebraic equations 
\begin{equation}
f(v_{j})=\sum_{k\neq j}^{M}\frac{2}{v_{k}-v_{j}},
\label{bae}
\end{equation}
where $f(v)$ is a rational function which can be expressed as
\begin{equation*}
f(v)=\sum_{i=0}^{k}c_i v^{i}+\sum_{m=0}^{s}\frac{1}{(v-a_{m})^{b_{m}}}.
\end{equation*}
Constructing the polynomial
\begin{equation*}
Q(z)=\prod_{k=1}^{M}(z-v_{k}),
\end{equation*}
it satisfies
\begin{equation*}
\frac{Q''(v_{j})}{Q'(v_{j})}=-\sum_{k\neq j}^{M}\frac{2}{v_{k}-v_{j}}.
\end{equation*}
Then Eq. (\ref{bae}) can be expressed in the polynomial form
\begin{equation*}
\left(\prod_{m=0}^{s}(v_{j}-a_{m})^{b_{m}}\right)Q''(v_{j})+\left(\prod_{m=0}^{s}(v_{j}-a_{m})^{b_{m}}\right)\left(\sum_{i=0}^{k}c_i v_{j}^{i}+\sum_{m=0}^{s}\frac{1}{(v_{j}-a_{m})^{b_{m}}}\right)Q'(v_{j})=0.
\end{equation*}
Using $Q(v_{j})=0$ we can write
\begin{equation}
A_{2}(z)Q''(z)+A_{1}(z)Q'(z)=A_{0}(z)Q(z),
\label{ode}
\end{equation}
where $A_{2}(z)$, $A_{1}(z)$ and $A_{0}(z)$ are  polynomials of order 
\begin{align*}
K_2&=\sum_{m=0}^{s}b_{m}, \\ 
K_1&=k+\sum_{m=0}^{s}b_{m}, \\
K_0&= k-1+\sum_{m=0}^{s}b_{m}
\end{align*}
 respectively. Specifically,
\begin{align*}
A_{2}(z)&=\prod_{m=0}^{s}(z-a_{m})^{b_{m}}, \\
A_{1}(z)&=\left(\sum_{i=0}^{k}c_iz^{i}+\sum_{m=0}^{s}\frac{1}{(z-a_{m})^{b_{m}}}\right)\prod_{n=0}^{s}(z-a_{n})^{b_{n}}.
\end{align*}
The polynomials $Q(z)$ and $A_{0}(z)$ can be constructed by inserting expansions 
\begin{equation}
Q(z)=\sum_{k=0}^{M}\alpha_{k}z^{k},\quad A_{0}(z)=\sum_{j=0}^{K_0}\beta_{j}z^{j}
\label{expans}
\end{equation} 
into (\ref{ode}), yielding a system of linear equations. We take $\alpha_{M}=1$ with the remaining 
$\alpha_{k}$ and $\beta_{j}$ to be determined numerically. The roots of the polynomial $Q(z)$ can be extracted once the $\alpha_{k}$ have been computed.   
We will follow the numerical procedure given in \cite{Mar12}. Related approaches are described in  \cite{Ler13,Van14,Pan11,Far11,El12,Guan12}. 
 
For the two models to be analysed below it is found that the roots associated with the ground state lie on the real line. In such an instance the discrete root density is computed from the numerical solution via 
\begin{equation*}
\tilde{\rho}(v_{j})=\frac{1}{(M-1)(v_{j+1}-v_{j})}, \qquad j=1,...,M-1
\end{equation*}
such that
\begin{equation}
  \sum_{j=1}^{M-1} \tilde{\rho}(v_{j})(v_{j+1}-v_{j})=1.
\label{disnorm}
\end{equation}
On the other hand, in the limit $M\rightarrow \infty$ a root density $\rho(v)$ with support on an interval $[\mathfrak{a},\mathfrak{b}]\subseteq {\mathbb R}$ can be introduced as a solution of the continuum limit of (\ref{bae}), viz. the singular integral equation
\begin{align}
\lim_{M\rightarrow\infty}\frac{f(v)}{M}=P\int_\mathfrak{a}^\mathfrak{b} \frac{2\rho(w)}{w-v} \,dw,
\label{sie}
\end{align}
where $P$ denotes the Cauchy principal value of the integral, subject to 
\begin{align}
\int_\mathfrak{a}^\mathfrak{b} \rho(w)\,dw=1
\label{connorm}
\end{align}
such that (\ref{connorm}) is the continuum analogue of (\ref{disnorm}).

In this paper we will compute both the discrete and continuum root densities for two models studied in \cite{Rub12}, which will be explicitly provided below. Moreover, it will be demonstrated how the root densities can be used to perform calculations which identify a quantum phase transition in each system. 

\section{Atomic-molecular Bose-Einstein condensate model}

The first Hamiltonian to be studied takes the following form
\begin{equation}
H=\mu \hat{N_{c}}+\Omega (\hat{a}^{\dagger}\hat{b}^{\dagger}\hat{c}+\hat{c}^{\dagger}\hat{a}\hat{b}),
\label{abc}
\end{equation}
where the operators $\{\hat{j},\hat{j}^{\dagger}|\hat{j}=\hat{a},\hat{b},\hat{c}\}$ are canonical bosonic creation and annihilation operators, and $\hat{N}_j=\hat{j}^\dagger \hat{j}$. The parameter $\mu$ governs the external potential and $\Omega$ is the amplitude for interconversion of atoms, associated with labels $a$ and $b$,  and molecules, associated with hte label $c$. The Hamiltonian (\ref{abc}) is a particular limit of a more general model for hetero-nuclear atomic-molecular Bose--Einstein condensates, introduced in \cite{Dun07,Zhou07,Zhou08}. The form (\ref{abc}) appeared many years ago in quantum optics \cite{Walls70}, and it is also the analogue of the homo-nuclear model studied by Vardi, Yurovsky, and Anglin \cite{Var01}. 

This system is integrable and exactly solvable. The Hamiltonian (\ref{abc}) commutes with the total number of particles $\hat{N}=\hat{N}_{a}+\hat{N}_{b}+2\hat{N}_{c}$ and the atomic imbalance $\hat{J}=\hat{N}_{a}-\hat{N}_{b}$.
We denote the eigenvalues of $\hat{N}$ and $\hat{J}$ by $N$ and $J$ respectively. The energy eigenvalues are given by \cite{Dun07}
\begin{equation}
E=-\Omega \sum_{j=1}^{M}v_{j},
\label{abcnrg}
\end{equation}
where the $v_j$ are roots of the associated Bethe Ansatz equations
\begin{equation}
\frac{J+1}{v_{j}}-v_{j}-\frac{\mu}{\Omega}=\sum_{k\neq j}^{M}\frac{2}{v_{k}-v_{j}}
\label{abcbae}
\end{equation}
with $M=(N-J)/2$ and $J=0,1,...,N$. We also introduce the fractional imbalance $k={J}/{N} \in [-1,1]$.

\subsection{Continuum limit approximation}

The Bethe Ansatz equations (\ref{abcbae}) in the continuum limit $M\rightarrow\infty$ take the form of a singular integral equation. For technical reasons it is most convenient to consider the integral form (\ref{sie}) as an approximation for the Bethe Ansatz equations (\ref{abcbae}) for large, but finite, $M$. This yields 
\begin{equation}
P\int_{\mathfrak{a}}^\mathfrak{b}\frac{2\rho(w)}{w-v}\,dw=\frac{1}{M}\left(\frac{J+1}{v}-v-\frac{\mu}{\Omega}                \right)
\label{abccontbae}
\end{equation}
such that $M$ appears explicitly as a variable. This approximation will allow us to determine the scaling properties of certain quantities as $M\rightarrow\infty$, which is necessary for an intermediate step in the calculations below. 

Next we adopt the following Ansatz for the root density
\begin{equation}
\rho(v)=\sqrt{(\mathfrak{b}-v)(v-\mathfrak{a})}\left(A+\frac{B}{v}\right)
\label{dens}
\end{equation}
with $A$ and $B$ some constants yet to be determined. Due to the branch cut in (\ref{dens}) the integral in the left-hand side of (\ref{abccontbae}) can be evaluated over a contour in the complex plane which encloses the interval $[\mathfrak{a},\mathfrak{b}]$. The contour integral can be evaluated by computing the residues at the origin and at the point at infinity. See Appendix B of \cite{Mar12} for further details. Performing these calculations produces
\begin{align*}
\int_{\mathfrak{a}}^\mathfrak{b}\rho(v)\,dv&= \frac{A\pi}{8}(\mathfrak{a}-\mathfrak{b})^{2}+\frac{B\pi}{2}(\sqrt{\mathfrak{a}}-\sqrt{\mathfrak{b}})^{2},  \\
P\int_{\mathfrak{a}}^\mathfrak{b}\frac{2\rho(w)}{w-v}\,dw&=A\pi (\mathfrak{a}+\mathfrak{b}-2 v)+2 B\pi \left(\frac{\sqrt{\mathfrak{ab}}}{v}-1\right).
\end{align*}
This leads to the following four equations for the parameters $A$, $B$, $\mathfrak{a}$ and $\mathfrak{b}$ in terms of $\mu$, $\Omega$ and $M$:
\begin{equation}
\begin{aligned}
1&=\frac{A\pi}{8}(\mathfrak{a}-\mathfrak{b})^{2}+\frac{B\pi}{2}(\sqrt{\mathfrak{a}}-\sqrt{\mathfrak{b}})^{2} , \\
J+1&=2\pi B M\sqrt{\mathfrak{ab}}, \\
1&=2\pi A M, \\
-\frac{\mu}{\Omega}&=\pi A (\mathfrak{a}+\mathfrak{b})M-2\pi B M.
\end{aligned}
\label{aandb}
\end{equation}
Rearranging the second and third equations to obtain
\begin{equation*}
A=\frac{1}{2\pi M}, \qquad  B=\frac{J+1}{2\pi M\sqrt{\mathfrak{ab}}}, 
\end{equation*}
and inserting in the two other equations, yields
\begin{align*}
 M&=\frac{1}{16}(\mathfrak{a}-\mathfrak{b})^{2}+\frac{J+1}{4\sqrt{\mathfrak{ab}}}(\sqrt{\mathfrak{a}}-\sqrt{\mathfrak{b}})^{2}, \\
 -\frac{\mu}{\Omega}&=\frac{\mathfrak{a}+\mathfrak{b}}{2}-\frac{J+1}{\sqrt{\mathfrak{ab}}}.
\end{align*}
Setting
\begin{align*}
\alpha=-\frac{\mu}{\Omega\sqrt{2N}},
\end{align*}
we can manipulate the above to obtain the following quartic equation for $\sqrt{\mathfrak{ab}}$:
\begin{equation}
 (\mathfrak{ab})^{2} + 2(1-\alpha^{2})N \mathfrak{ab} - 4(J+1)\alpha \sqrt{2N}  \sqrt{\mathfrak{ab}}-3 (J+1)^{2}  =0.
\end{equation}
Assuming $J=O(N^0)$ we have the following asymptotics:
\begin{align}
\mathfrak{ab}\sim\begin{cases} \displaystyle 
2(\alpha^2-1)N,  &\alpha>1,\\
2^{{5}/{3}}(J+1)^{{2}/{3}}N^{{1}/{3}},   &\alpha=1,  \\
 \displaystyle \left(\frac{J+1}{f}\right)^{2}  N^{-1},  & \alpha<1,
\end{cases}
\label{asympt}
\end{align}
where
$$f=\frac{2(1-\alpha^2)}{2\sqrt{2}\alpha+\sqrt{2\alpha^2+6}}. $$
Conversely, for $J=O(N)$ such that $k=J/N\neq 0$, we have $\mathfrak{ab}=O(N)$ for all $\alpha$.

\begin{figure}
\centering
\subfloat{\includegraphics[width=6.05cm]{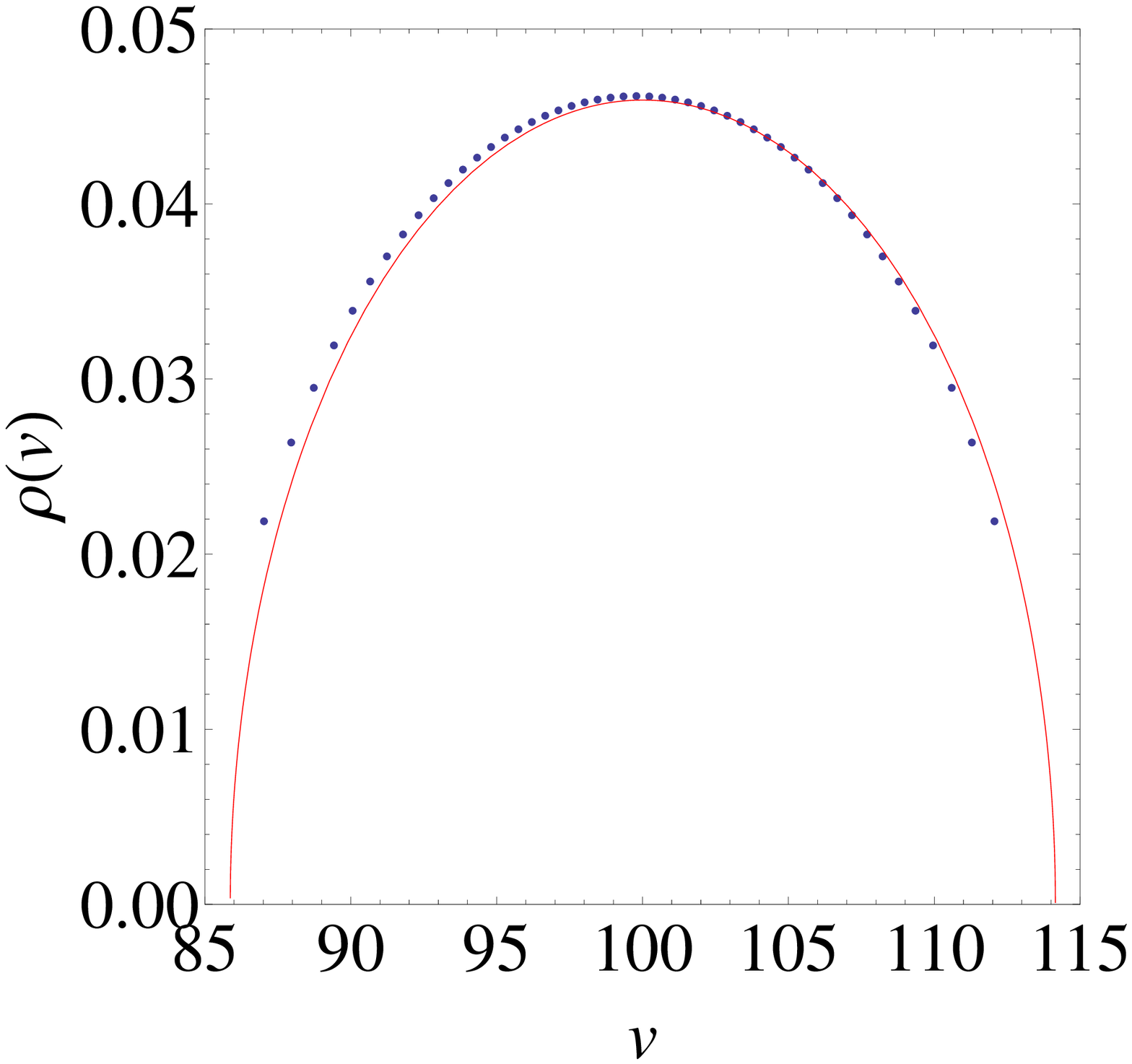}}
\subfloat{\includegraphics[width=6cm]{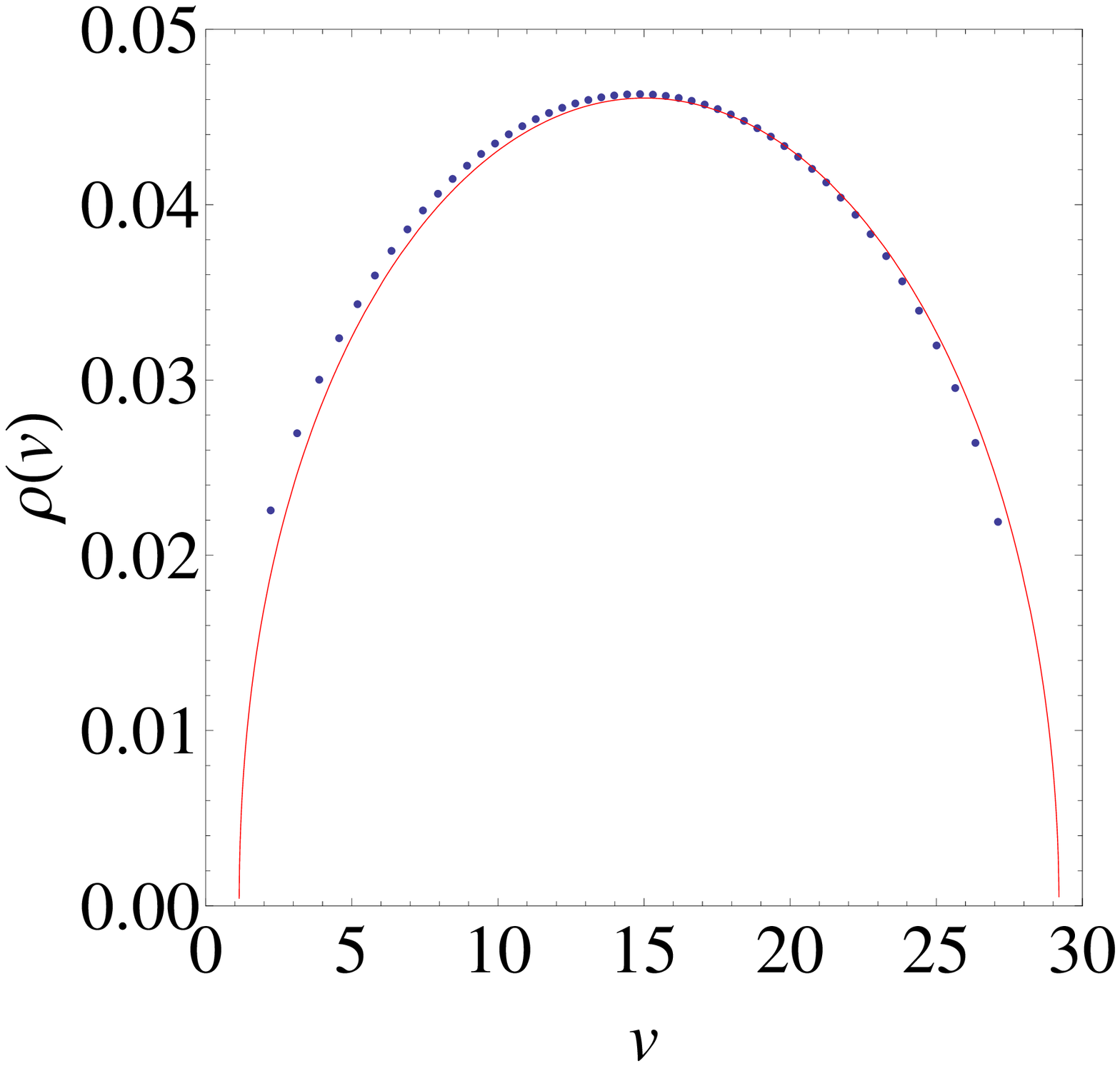}}
\caption{Atomic-molecular Bose--Einstein condensate model: Ground-state roots densities for $M=50,\,J=0$ and $\Omega=1$. The discrete root density is depicted by points and the solid line is the continuum limit approximation for (a) $\mu =-100$ ($\alpha\approx 7.07 $), (b) $\mu =-15$ ($\alpha\approx  1.06$). In both cases the continuum limit approximation vanishes at $\mathfrak{a}>0$.}
\label{fig1}
\end{figure}

\begin{figure}
\centering
\subfloat{\includegraphics[width=6cm]{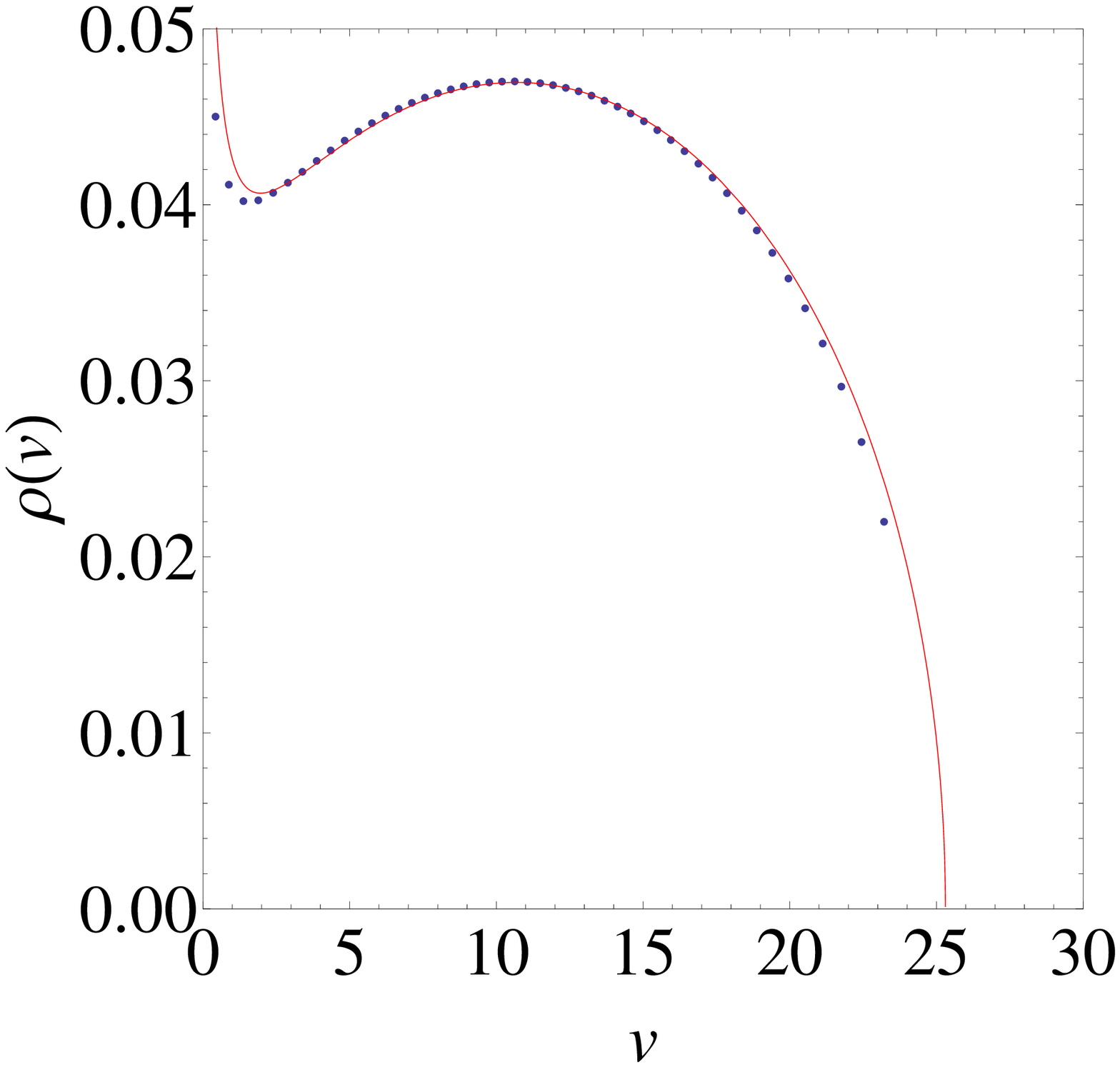}}
\subfloat{\includegraphics[width=6cm]{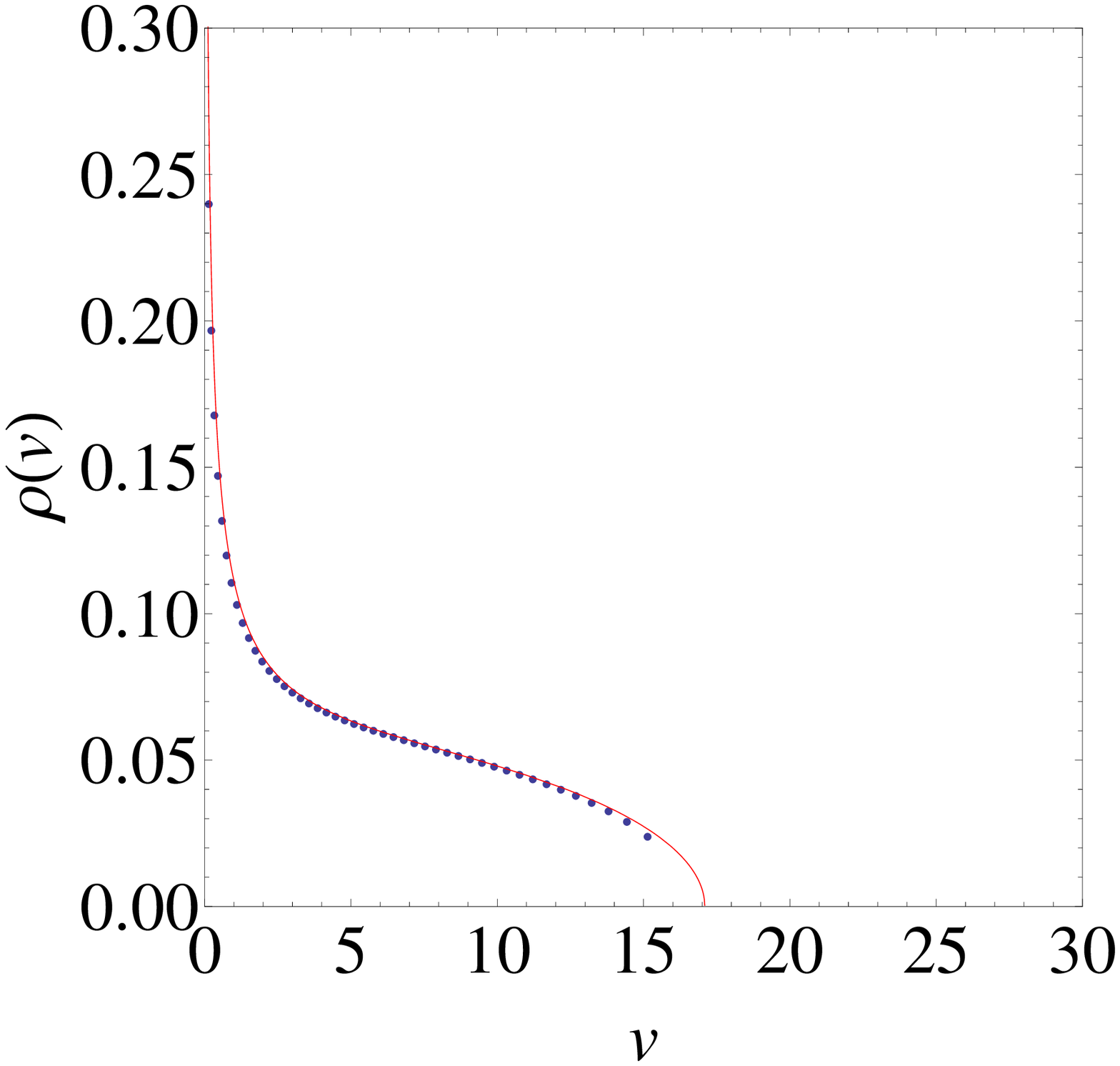}}
\caption{Atomic-molecular Bose--Einstein condensate model: Ground-state roots densities for $M=50,\,J=0$ and $\Omega=1$. The discrete root density is depicted by points and the solid line is the continuum limit approximation for (a) $\mu =-11$ ($\alpha\approx 0.78 $), (b) $\mu =-1$ ($\alpha\approx  0.07$). In both cases the continuum limit approximation diverges at $\mathfrak{a}=0$.}
\label{fig2}
\end{figure}

In \cite{Dun07} the coupling $\alpha=1$ was identified as a quantum phase transition point when $k=0$, but it was also found that there is no transition for $k\neq 0$. Here, the quantum phase transition manifests as a change in the scaling of $\mathfrak{ab}$ when $k=0$, while there is no such change for $k\neq 0$. For $k=0$ there is a distinct qualitative change in the ground-state root density upon crossing $\alpha=1$. This is illustrated in Fig. \ref{fig1}, where both the discrete root density and continuum approximation are plotted for $M=50,\,J=0$ and particular values of $\alpha>1$.  As the coupling parameter $\alpha$ decreases the quantity $\mathfrak{a}$, the minimum endpoint of the support for the root density, moves towards zero. For comparison, analogous densities are plotted in Fig. \ref{fig2} for $M=50,\,J=0$ and particular values of $\alpha<1$. The main qualitative difference between the two figures is the behaviour of the root density at $\mathfrak{a}$. In Fig. \ref{fig1}, the continuum limit approximation vanishes at $\mathfrak{a}>0$. In Fig. \ref{fig2}, the continuum limit approximation diverges at $\mathfrak{a}=0$. Also note the change in the vertical scale of the second panel in Fig. \ref{fig2}.

\subsection{Ground-state energy and molecular fraction expectation value}

Having established that a sudden change occurs in the ground-state root density upon crossing the point $\alpha=1$, we next demonstrate how this manifests in certain physical quantities. First we consider the ground-state energy. From (\ref{abcnrg},\ref{dens}), the continuum limit approximation becomes
\begin{align}
E&=-\Omega \int_{\mathfrak{a}}^{\mathfrak{b}}\rho(v) v \,dv \nonumber \\
&= \mu \left(\frac{N+1}{2}+\frac{(J+1)^{2}}{2\mathfrak{ab}}\right)-\frac{ (J+1)\mu }{2 \alpha \sqrt{2N}\sqrt{\mathfrak{ab}}}\left( \mathfrak{ab} - \frac{(J+1)^{2}}{\mathfrak{ab}}\right),
\label{inter}
\end{align}
where the expression above has been obtained via contour integral techniques and simplified using the Eqs. (\ref{aandb}). Use of (\ref{asympt}) then yields the leading order behaviour 
\begin{align}
E\sim\begin{cases} \displaystyle  
\frac{\mu N}{2}, & \alpha \geq 1, \\
\\
\displaystyle \frac{\mu N}{2}\left(1+f^2+\frac{1}{\sqrt{2}\alpha}f^3  \right)  , & \alpha\leq 1.
\end{cases}
\label{abccontnrg}
\end{align}  
It is somewhat surprising that the above result is independent of $J$, despite (\ref{asympt},\ref{inter}) being $J$-dependent.

\begin{table}
\footnotesize\rm
\begin{tabular*}{\textwidth}{@{}l*{15}{@{\extracolsep{0pt plus12pt}}l}}
\br
 $\mu$ & $E_{cl}$    & $E_{num}$ & $\Delta E$ &  \%  error \\
\br
 -100 & -5000.000  & -5000.500 & 0.500 & 0.001 \\
 -15  & -750.000  & -754.718 &  4.718   & 0.625 \\
 -11  &  -566.498 &  -570.397  & 3.899  & 0.684  \\
  -1  & -289.531  & -290.763 & 1.232 &  0.423 \\  
\mr
\end{tabular*}
\caption{Atomic-molecular Bose--Einstein condensate model: Ground-state energy for $N=100$, $J=0$ and $\Omega=1$. For each value of $\mu$, $E_{cl}$ denotes the value obtained from the continuum limit approximation (\ref{abccontnrg}), while $E_{num}$ is obtained from (\ref{abcnrg}) and the numerical solution of the Bethe Ansatz equations (\ref{abcbae}). The final columns show the difference and the relative percentage error respectively. The quantum phase transition point is $\mu_c\approx -14.14$. }
\end{table}

Table 1 compares the ground-state energy from the continuum approximation against results obtained by numerically solving the Bethe Ansatz equations (\ref{abcbae}) and using (\ref{abcnrg}). The agreement is excellent. It is anticipated that the continuum limit approximation becomes exact as $N\rightarrow\infty$. However taking this limit from the outset is problematic. In particular this can be seen through the $N$-dependence of (\ref{asympt}), which is required to compute (\ref{abccontnrg}) via (\ref{inter}). 

To conclude this discussion, we show how this approach enables the characterisation of  the quantum phase transition at $\alpha=1$ via an order parameter. Recall that the Hellmann-Feynmann theorem can be stated as
\begin{equation*}
\left\langle \frac{\partial H}{\partial \lambda} \right\rangle = \frac{\partial E}{\partial \lambda},
\end{equation*}
where in general $H$ is a Hamiltonian depending on a coupling parameter $\lambda$, and the expectation value is with respect to an eigenstate of energy $E$. Defining  
$$\mathcal{O}= \frac{2\left\langle N_{c} \right\rangle}{N},$$ which is the molecular fraction expectation value, it follows for the Hamiltonian (\ref{abc}) that
\begin{align*}
\mathcal{O}=\frac{2}{N}\frac{\partial E}{\partial \mu}.
\end{align*}
From (\ref{abccontnrg}) we obtain for the ground-state molecular fraction expectation value
\begin{align}
\mathcal{O}\sim\begin{cases} \displaystyle  
1, & \alpha \geq 1, \\
\displaystyle 1+f^2+\left(2\alpha f+\frac{3}{\sqrt{2}}f^2  \right)f'  , & \alpha\leq 1,
\end{cases}
\label{mfev}
\end{align}  
where $f'$ denotes the derivative of $f$ with respect to $\alpha$. 
This calculation shows that the quantum phase transition point 
$\alpha=1$ separates a pure molecular phase and a mixed atomic-molecular phase.
\hspace{-10cm}
\begin{figure}
\centering
\subfloat{\includegraphics[width=6cm]{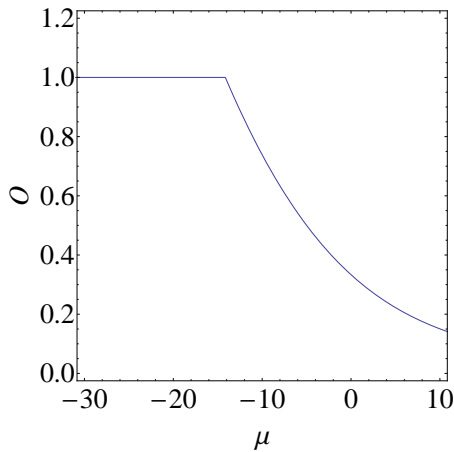}}
\caption{Atomic-molecular Bose--Einstein condensate model: Molecular fraction expectation value $\mathcal{O}$, as given by the continuum approximation result (\ref{mfev}), as a function of the coupling parameter $\mu$ for $N=100$, $J=0$ and $\Omega=1$. The quantum phase transition point is $\mu_c\approx -14.14$. Compare with Fig. 7 of \cite{Dun07} which has similar qualitative features. }
\label{fig3}
\end{figure}

\section{Symmetric two-site Bose--Hubbard model}

The Hamiltonian of the symmetric two-site Bose--Hubbard model is given by \cite{Mil97,Cir98,Leg01,Zhou03,Pan05,Kohl02,Perez10,Polls10,Buo12,Sim12,Jez13,Gra14,Sak14}
\begin{equation}
H=\frac{k}{8}(\hat{N}_{1}-\hat{N}_{2})^{2}-\frac{\mathcal{E}}{2}(\hat{a}_{1}^{\dagger}\hat{a}_{2}+\hat{a}_{2}^{\dagger}\hat{a}_{1}),
\label{aaham}
\end{equation}
where for $i,j=1,2$
\begin{equation*}
[\hat{a}_{i},\hat{a}_{j}^{\dagger}]=\delta_{ij},\quad [\hat{a}_{i},\hat{a}_{j}]=[\hat{a}_{i}^{\dagger},\hat{a}_{j}^{\dagger}]=0,
\end{equation*}
and $\hat{N}_{j}=\hat{a}_{j}^{\dagger}\hat{a}_{j}$. Setting $\hat{N}=\hat{N}_{1}+\hat{N}_{2}$, it can be verified that $[H,\,\hat{N}]=0$. We denote the eigenvalues of $\hat{N}$ by $N$. Note that because (\ref{aaham}) is invariant under the interchange of labels 1 and 2, the subspaces of symmetric and antisymmetric states are invariant under the action of (\ref{aaham}). 

Previous studies \cite{Cir98,Kohl02,Pan05} have identified a quantum phase transition in the attractive regime $k<0.$ Setting
\begin{align}
\lambda=-\frac{kN}{2\mathcal{E}},
\label{lambda}
\end{align}
the transition takes place at $\lambda=1$. This phenomenon was examined in \cite{Rub12} with attention to the nature of the ground-state roots of the following Bethe Ansatz equations 
\begin{equation}
\frac{\mathcal{E} v_{j}^{2}+k(1-N)v_{j}-\mathcal{E}}{k v_{j}^{2}}=\sum_{k \neq j}^{N}\frac{2}{v_{k}-v_{j}}.
\label{oldbae}
\end{equation}
We prefer to use an alternative form, which first appeared in \cite{Eno93}. For simplicity we restrict to the case where $N$ is even. The alternative Bethe Ansatz equations read
\begin{equation}
\frac{2\mathcal{E}}{k} +\frac{2\mu}{v_{j}-1}+\frac{2\mu}{v_{j}+1}=\sum_{k \neq j}^{M}\frac{2}{v_{k}-v_{j}},
\label{aabae}
\end{equation}
such that the associated energy is given by
\begin{equation}
E=\frac{kN^{2}}{8}+\mathcal{E} \sum_{j=1}^{M}v_{j},
\label{aanrg}
\end{equation}
the total particle number is
$$ N=2M + 4\mu -1,$$
and $\mu=1/4$ for symmetric states and $\mu=3/4$ for antisymmetric states. It can be checked that the ground state lies in the symmetric subspace of the full space of states, e.g. see \cite{Cir98}, in which case $N=2M$. 

The equivalence of the two forms of Bethe Ansatz equations (\ref{oldbae}) and (\ref{aabae}) was established in \cite{Lin09}\footnote{There is a typographical error in the energy expression (18) of Ref. \cite{Lin09}}.
The advantage of using the form (\ref{aabae}) is that the ground-state roots are real-valued and lie in the interval $[-1,\,1]$, which was deduced by numerical solution of the Bethe Ansatz equations (\ref{aabae}) using the techniques in \cite{Mar12}.

\subsection{Continuum limit approximation}

Adopting the procedure of the previous example we consider the integral form (\ref{sie}) as an approximation of (\ref{aabae}) for large, but finite, $M$:
\begin{equation}
\frac{4\alpha}{M} +\frac{1}{M(v-1)}+\frac{1}{M(v+1)}= P\int_{\mathfrak{a}}^{\mathfrak{b}}\frac{4\rho(w)}{w-v}\,dw,
\label{conaabae}
\end{equation}
where $-1\leq\mathfrak{a}< \mathfrak{b}\leq 1$ and we have set $\rho=1/4$. Taking the density to have the form
\begin{equation}
\rho(v)=\sqrt{(\mathfrak{b}-v)(v-\mathfrak{a})}\left(\frac{A}{v-1}+\frac{B}{v+1}\right)
\label{aarho}
\end{equation}
it follows from (\ref{conaabae}) that 
\begin{align*}
&\frac{4\alpha}{M}  +\frac{1}{M(v-1)}+\frac{1}{M(v+1)} \\
&\qquad = -4 \pi A\left( \frac{\sqrt{ (1-\mathfrak{a})(1-\mathfrak{b})}}{v-1}+1\right)+ 4  \pi B\left( \frac{\sqrt{ (1+ \mathfrak{a})(1+\mathfrak{b})}}{v-1}-1\right),
\end{align*}
while the normalisation condition (\ref{connorm}) gives 
\begin{equation*}
\frac{\pi A}{2}(\mathfrak{a}+\mathfrak{b}-2 +2 \sqrt{(1-\mathfrak{a})(1-\mathfrak{b})})+\frac{\pi B}{2}(\mathfrak{a}+\mathfrak{b}+2 -2 \sqrt{(1+\mathfrak{a})(1+\mathfrak{b})})=1.
\end{equation*}
Setting
\begin{align*}
\mathfrak{c}&=\sqrt{(1-\mathfrak{a})(1-\mathfrak{b})}, \\
\mathfrak{d}&=\sqrt{(1+\mathfrak{a})(1+\mathfrak{b})}
\end{align*}
it is deduced that 
\begin{align*}
A&=-\frac{1}{8\pi N\mathfrak{c}}, \\
B&=\frac{1}{8\pi N \mathfrak{d}}, \\
A+B&=\frac{1}{\pi\lambda}.
\end{align*}
Eliminating $A$ and $B$ then leads to the equations
\begin{align*}
\lambda(\mathfrak{d}^{-1}-\mathfrak{c}^{-1})&={8N},  \\
2N(\mathfrak{d}^{2}-\mathfrak{c}^{2})+\lambda(\mathfrak{d}^{-1} +\mathfrak{c}^{-1})&=4\lambda(2N+1).
\end{align*}
The leading order solution valid for $\lambda>1$ is
\begin{align*}
\mathfrak{c} \sim \frac{1}{4(1-\lambda^{-1})N}, \\
\mathfrak{d} \sim \frac{1}{4(1+\lambda^{-1})N},
\end{align*}
which shows that $\mathfrak{a}\sim -1$ and $\mathfrak{b}\sim 1$. For $\lambda<1$
\begin{align*}
\mathfrak{c} &\sim {2\sqrt{ 1-\lambda}},   \\
\mathfrak{d} &\sim \frac{\lambda}{8N}, 
\end{align*}
yielding
\begin{align*}
\mathfrak{a}&\sim -1, \\
\mathfrak{b}&\sim 2{\lambda} -1. 
\end{align*}

\begin{figure}
\centering
\subfloat{\includegraphics[width=6.1cm]{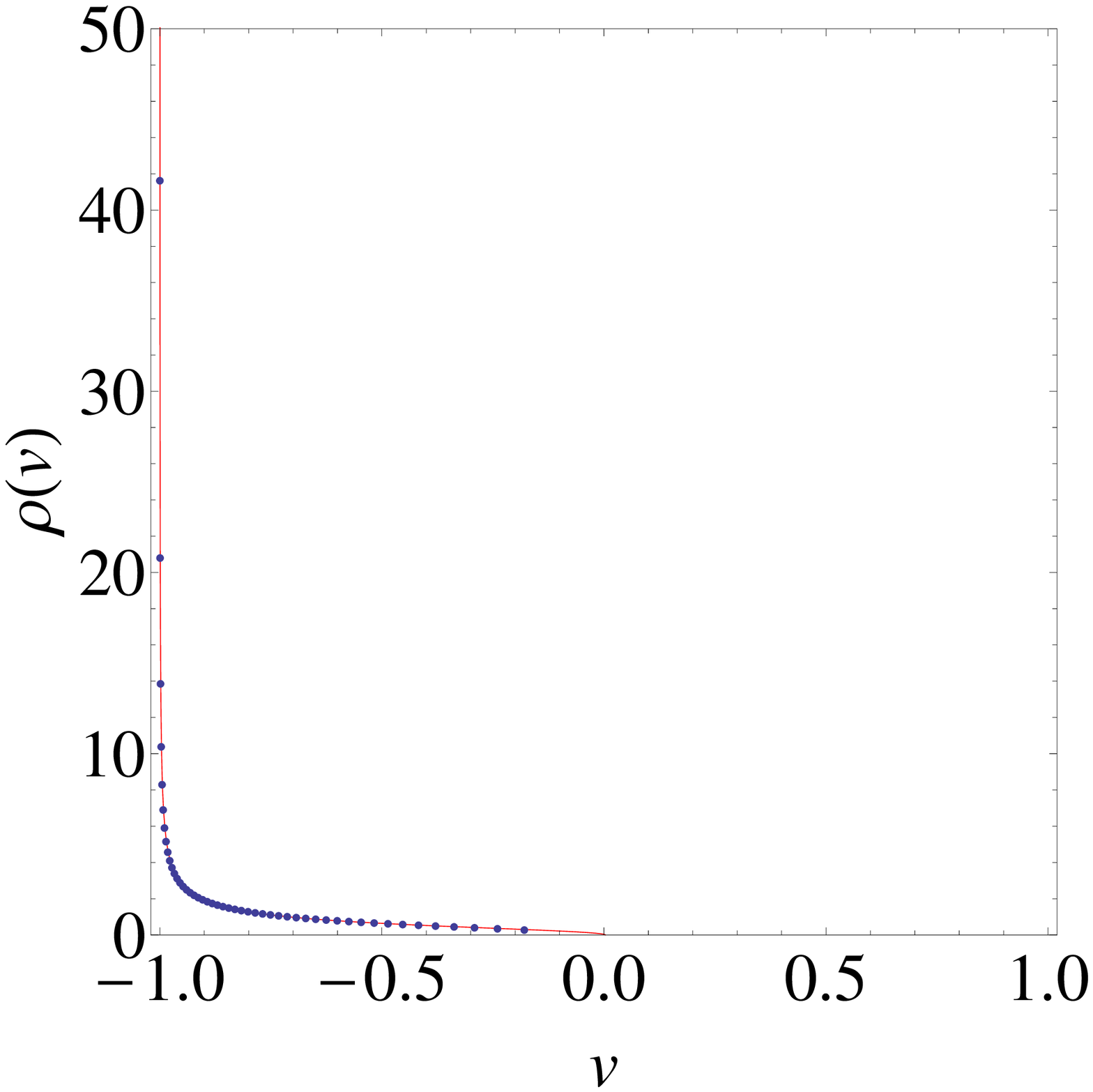}}
\subfloat{\includegraphics[width=6cm]{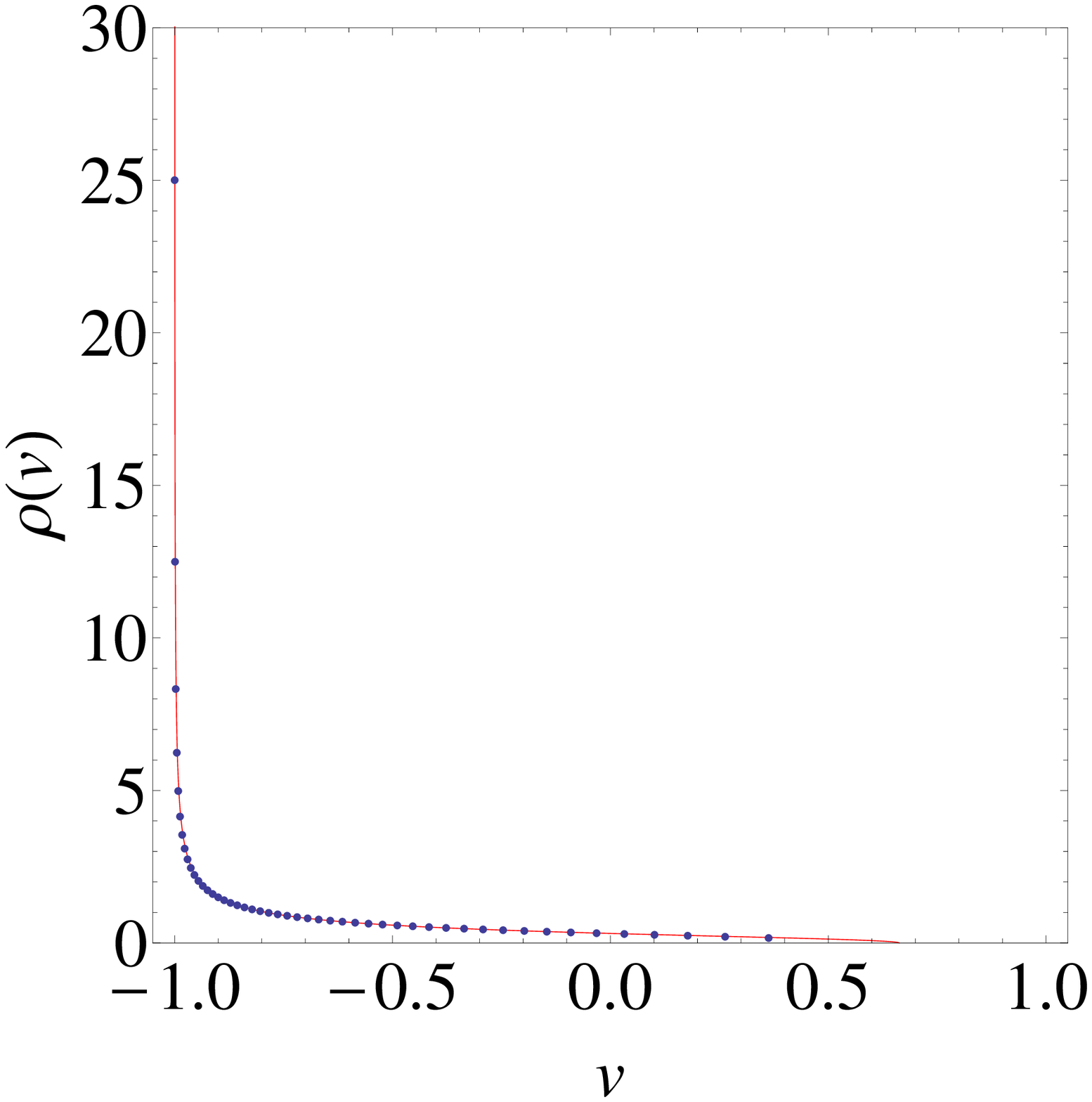}}
\caption{Symmetric two-site Bose--Hubbard model:
Ground-state roots densities for $M=50$ ($N=100$) and $\mathcal{E}=1$. The discrete root density is depicted by points and the solid line is the continuum limit approximation for (a) $k =-1/100$ ($\lambda=1/2 $), (b) $k =-1/60$ ($\lambda=5/6$). Here, the continuum limit approximation vanishes at $\mathfrak{b}<1$.}
\label{fig4}
\end{figure}

\begin{figure}
\centering
\subfloat{\includegraphics[width=6cm]{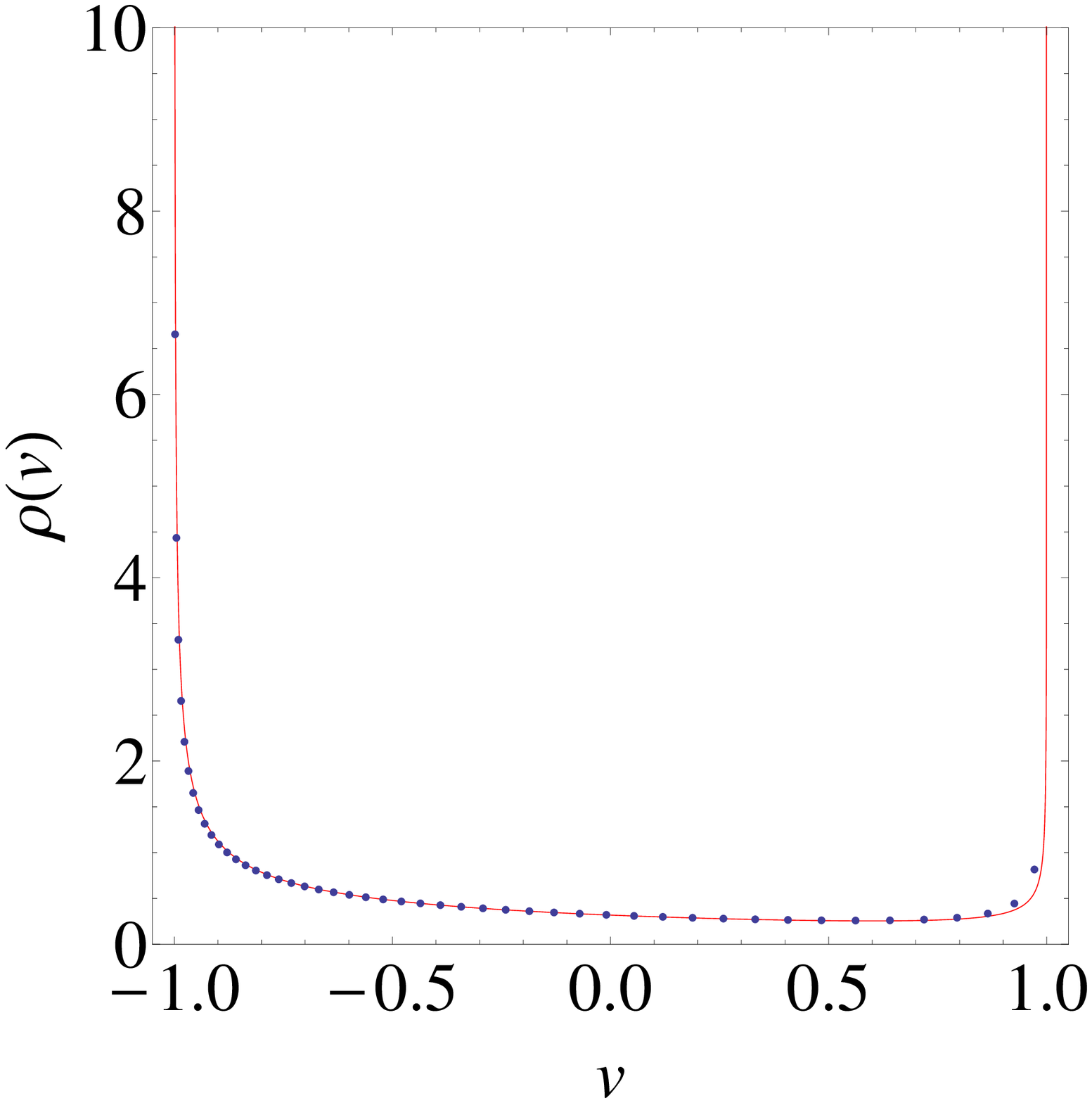}}
\subfloat{\includegraphics[width=6.1cm]{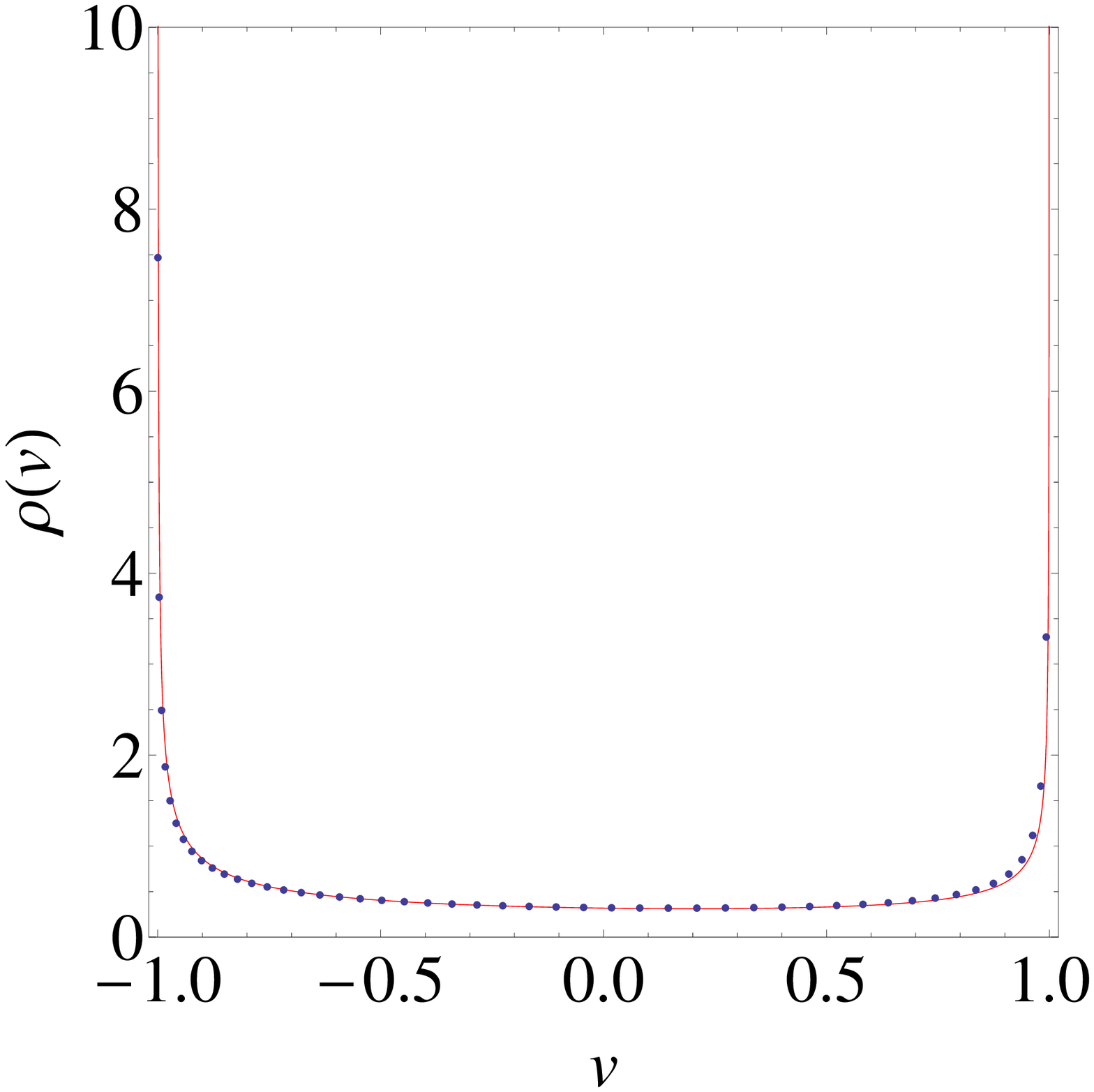}}
\caption{Symmetric two-site Bose--Hubbard model: 
Ground-state roots densities for $M=50$ ($N=100$) and $\mathcal{E}=1$. The discrete root density is depicted by points and the solid line is the continuum limit approximation for (a) $k =-1/30$ ($\lambda=5/3  $), (b) $k =-1/10$ ($\lambda =5$). In both cases the continuum limit approximation diverges at $\mathfrak{b}=1$.}
\label{fig5}
\end{figure}

In Fig. \ref{fig4}, both the discrete root density and continuum approximation are plotted for $M=50$, corresponding to $N=100$, and particular values of $\lambda<1$.  As the coupling parameter $\lambda$ increases the quantity $\mathfrak{b}$, the maximum endpoint of the support for the root density, moves towards 1. For comparison, analogous densities are plotted in Fig. \ref{fig5} for $M=50$ and particular values of $\lambda>1$. The main qualitative difference between the two figures is the behaviour of the root density at $\mathfrak{b}$. In Fig. \ref{fig4}, the continuum limit approximation vanishes at $\mathfrak{b}<1$. In Fig. \ref{fig5}, the continuum limit approximation diverges at $\mathfrak{b}=1$. In both cases the root density is divergent at $\mathfrak{a}=-1$.

\subsection{Ground-state energy and correlation functions}

From (\ref{aanrg},\ref{aarho}) the continuum limit approximation for the ground-state energy reads 
\begin{align*}
E&=\frac{kN^2}{8}+\frac{\mathcal{E} N}{2}\int_{\mathfrak{a}}^{\mathfrak{b}} v \rho(v)\, dv \\ 
&=\frac{kN^2}{8}\left(1-\frac{2}{\lambda}\right)\int_{\mathfrak{a}}^{\mathfrak{b}} v \rho(v)\, dv\\
&=\frac{kN^2}{8}\left(1-\frac{2}{\lambda}\right)\int_{\mathfrak{a}}^{\mathfrak{b}} dv \sqrt{(\mathfrak{b}-v)(v-\mathfrak{a})} 
\left( \frac{Av}{v-1} + \frac{Bv}{v+1}\right).  
\end{align*}
Using the results of the previous subsection leads to the simple leading order result for the ground-state energy:
\begin{align}
E\sim\begin{cases} \displaystyle  
-\frac{\mathcal{E} N}{2}, & \lambda \leq 1, \\
\\
\displaystyle -\frac{\mathcal{E}N}{4}\left(\lambda+\lambda^{-1}  \right)  , & \lambda\leq 1.
\end{cases}
\label{aacontnrg}
\end{align}

\begin{table}
\footnotesize\rm
\begin{tabular*}{\textwidth}{@{}l*{15}{@{\extracolsep{0pt plus12pt}}l}}
\br
 $k$ & $E_{cl}$    & $E_{num}$ & $\Delta E$  & \%   error \\
\br
  -1/100  &  -50.000     &   -50.146     &  0.146     &  0.291   \\
   -1/60  & -50.000  &  -50.292  & 0.292   & 0.581   \\
   -1/30  &  -56.667 & -56.836 &   0.169   & 0.297  \\
   -1/10  & -130.000  & -130.051 &  0.051  & 0.039  \\
\mr
\end{tabular*}
\caption{Symmetric two-site Bose--Hubbard model: 
Ground-state energy for $N=100$ and $\mathcal{E}=1$. For each value of $k$, $E_{cl}$ denotes the value obtained from the continuum limit approximation (\ref{aacontnrg}), while $E_{num}$ is obtained from (\ref{aanrg}) and the numerical solution of the Bethe Ansatz equations (\ref{aabae}). The final columns show the difference and the relative percentage error respectively. The quantum phase transition point is $k_c=-1/50$. }
\end{table}

As before, we appeal to the Hellmann-Feynman theorem to compute ground-state correlation functions. Following \cite{Zhou03} we define the coherence correlator to be given by 
\begin{align*}
\theta &=\frac{1}{N} \left\langle  \hat{a}_{1}^{\dagger}\hat{a}_{2}+\hat{a}_{2}^{\dagger}\hat{a}_{1} \right\rangle \\
&=\frac{2}{N}\frac{\partial E}{\partial \mathcal{E}}
\end{align*}
and the imbalance fluctuation as 
\begin{align*}
\chi &=\frac{1}{N^{2}} \left\langle  (\hat{N}_{1}-\hat{N}_{2})^{2} \right\rangle \\
&=\frac{4}{N^2}\frac{\partial E}{\partial k}.
\end{align*}
From (\ref{aacontnrg}) these are found to be given by 
\begin{align*}
    \theta    &= \begin{cases}
                        1  \phantom{ - \lambda^{-2}} &   \lambda \leq 1  \\
                        \lambda^{-1}  &   \lambda \geq 1
                    \end{cases}  \\
    \chi    &= \begin{cases}
                        0    &  \lambda \leq 1  \\
                      1-  \lambda^{-2}  & \lambda \geq 1
                    \end{cases}
\end{align*}
The above formulae complement the asymptotic results of \cite{Zhou03}, which were derived for the repulsive case $k>0$.  

Finally, we can also use the above results to associate the quantum phase transition point $\lambda=1$ \cite{Polls10,Sak14,Zhu14} with the onset of condensate fragmentation. Following \cite{Polls10,Sak14,Zhu14}, 
denoting the ground state by $|\psi\rangle$ consider the one-body density matrix 
\begin{equation*}
\rho^{(1)}= \frac{1}{N}  \begin{pmatrix} 
  \langle \psi |\hat{a_{1}}^{\dagger} \hat{a_{1}} |\psi\rangle    
  &  \langle \psi |\hat{a_{1}}^{\dagger} \hat{a_{2}} |\psi\rangle   \\ 
  \langle \psi |\hat{a_{2}}^{\dagger}   \hat{a_{1}}|\psi\rangle 
  & \langle \psi |\hat{a_{2}}^{\dagger} \hat{a_{2}} |\psi\rangle
\end{pmatrix} .
\end{equation*}
The system is said to be unfragmented if the eigenvalues of $\rho^{(1)}$ are 0 and 1,  otherwise the system is said to be fragmented. Exploiting the symmetry of the Hamiltonian (\ref{aaham}) upon interchange of the labels 1 and 2, it is found that for $\lambda \leq 1$
\begin{equation*}
\rho^{(1)}=    \frac{1}{2}\begin{pmatrix} 
    {1}  &  {1}  \\ 
  {1} & {1} 
\end{pmatrix} 
\end{equation*}
with eigenvalues 0 and 1, while for $\lambda  \geq 1$
\begin{equation*}
\rho^{(1)}=    \frac{1}{2}\begin{pmatrix} 
    {1}  &  \lambda^{-1}  \\ 
  \lambda^{-1} & {1} 
\end{pmatrix} 
\end{equation*}
with eigenvalues $\displaystyle \frac{1}{2} \pm   \frac{1}{2\lambda}$. Thus, the phase transition point 
$\lambda=1$ separates fragmented and unfragmented phases. 

\section{Conclusion}

In this work we have re-examined the studies conducted in \cite{Rub12} for an atomic-molecular Bose--Einstein condensate model and the symmetric two-site Bose--Hubbard model. By calculation of the ground-state Bethe root density in the limit of infinite number of roots, we obtain analytic expressions for the ground-state energy which shows excellent agreement with numerical calculations. This in turn allows for the calculation of correlation functions through use of the Hellmann-Feynman theorem. These techniques are not specific to the two models considered here, but have wider applicability to other systems such as  
\cite{Dun10,Rom10,Ler11,Mar12,Ler13,Mar13,Van14} as mentioned in the Introduction. For new applications, we specifically identify the wide scope to apply these methods to generalisations of the two-site Bose--Hubbard model to cases which include non-linear tunneling \cite{Zhu14,Lia09,Ton13} and multi-level systems \cite{San13}.  
\ack
%
The research of J.\ L.\ is supported by the Australian Research Council through Discovery Project DP110101414, and I.\ M.\ is supported by Discovery Early Career Researcher Award DE130101067. We thank Angela Foerster for insightful advice, and Inna Lukyanenko for her astute comments.


\end{document}